# Tuning friction atom-by-atom in an ion-crystal simulator

Alexei Bylinskii[1†], Dorian Gangloff[1†], Vladan Vuletić[1]*

[1] Dept. of Physics, Massachusetts Institute of Technology, 77 Massachusetts Avenue, Cambridge, MA 02139, USA.

* Correspondence to: vuletic@mit.edu

† These authors contributed equally.

**Friction between ordered, atomically smooth surfaces at the nanoscale (nanofriction) is often governed by stick-slip processes. To test long-standing atomistic models of such processes, we implement a synthetic nanofriction interface between a laser-cooled Coulomb crystal of individually addressable ions as the moving object, and a periodic light-field potential as the substrate. We show that stick-slip friction can be tuned from maximal to nearly frictionless via arrangement of the ions relative to the substrate. By varying the ion number, we also show that this strong dependence of friction on the structural mismatch, as predicted by many-particle models, already emerges at the level of two or three atoms. This model system enables a microscopic and systematic investigation of friction, potentially even into the quantum many-body regime.**

Stick-slip friction is a non-linear phenomenon in which two surfaces stick to each other owing to a static friction force and accumulate potential energy under increasing applied shear force, then slip suddenly. As the released energy is dissipated, the surfaces stick again and the process repeats (*1*). This phenomenon occurs on length scales ranging from nanometers (biological molecules and atomic contacts (*1–3*)) to the kilometer scales of earthquakes (*4*). Interestingly, at the nanoscale, lattice mismatch between surfaces can cancel the sticking forces, resulting in continuous and almost frictionless sliding termed superlubricity (*5*). Despite their fundamental and technological importance, stick-slip and superlubricity are not fully understood, because of the difficulty of probing an interface with microscopic resolution and control.

The simplest atomistic friction model is the single-particle stick-slip model by Prandtl and Tomlinson (PT) (*6, 7*). The particle, held in a harmonic potential of an elastic object crystal, is driven across a sinusoidal potential of a rigid substrate crystal. This one-particle model, however, fails to capture the effects of structural mismatch between the crystal surfaces. The Frenkel-Kontorova (FK) model (*8, 9*) instead treats the object as an infinite array of atoms joined by springs. This model is governed by the commensurability of the unperturbed array and the substrate, and exhibits non-trivial kink dynamics (*8*), the pinned-to-sliding Aubry phase transition (*10*), and the related superlubricity (*5*).

Tools based on atomic force microscopy (*11*) can measure atomic-scale slips between surfaces comprising down to a few atoms (*12–14*). This has enabled the observation of superlubricity by varying the normal load (*15*) or the relative orientation of crystal lattices forming the interface (*16, 17*). Most observations in these systems can be qualitatively explained via variants of the PT or FK models, but without direct access to microscopic dynamics. Kink propagation dynamics, however, was observed in a macroscopic friction simulator with colloidal polystyrene beads in an optical lattice (*18*).

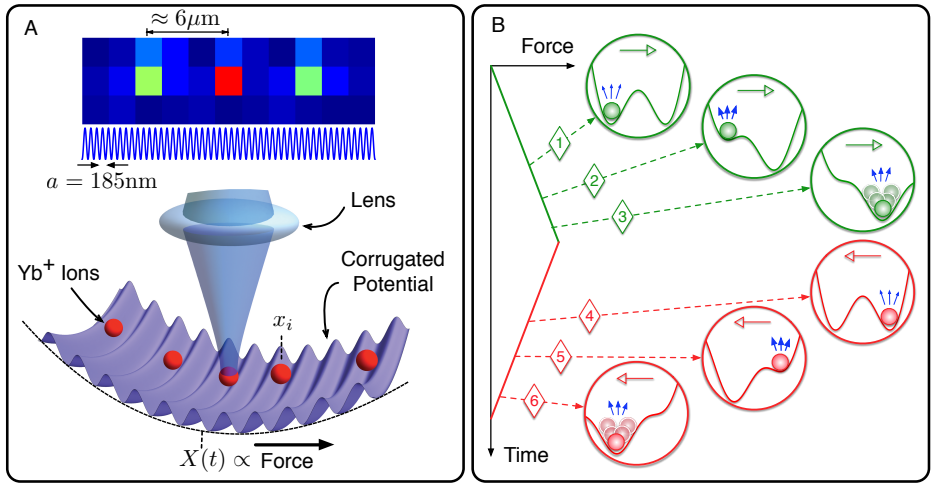

**Fig. 1**. **Ion-crystal simulator of stick-slip friction:** **(A)** Synthetic nanofriction interface between a Coulomb crystal of $^{174}$Yb$^+$ ions and an optical lattice, with single-ion-resolving microscope. The typical ion spacing is 6 μm, and the lattice period $a = 185$ nm. In the bottom illustration of the corrugated potential, the lattice period and the corrugation are strongly exaggerated. **(B)** Stick-slip results from bistability, illustrated here for a single ion. We linearly ramp a shear force causing the ion to jump between the minima, and we extract its position from its fluorescence, proportional to the lattice potential energy: (#1) ion initialized in the left site; (#2) the applied force pushes the ion up the lattice potential, eventually causing the slip; (#3) immediately after the slip, the ion is optically recooled and localizes to the right site; (#4), (#5), (#6) the force ramp reverses and the ion sticks at the right site before slipping back to the left. Slips are identified by maxima in the ion's fluorescence.

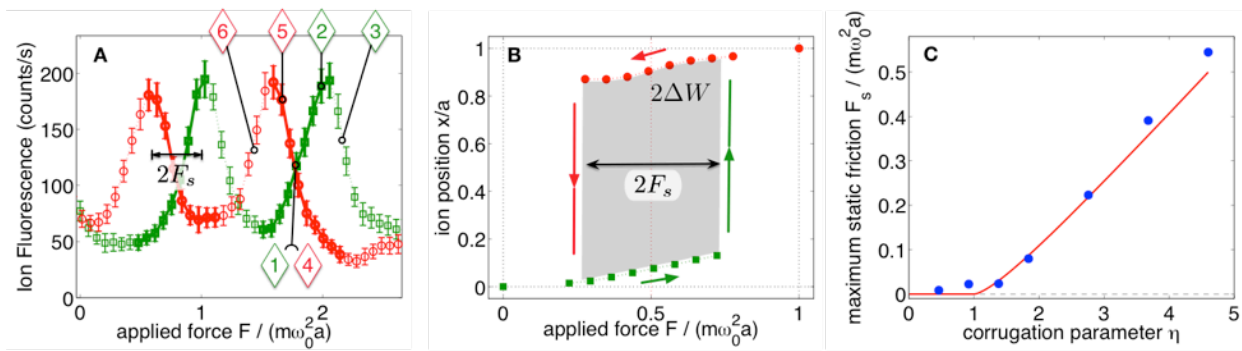

**Fig. 2. Measured stick-slip hysteresis cycle of a single ion.** **(A)** Fluorescence versus applied force during the forward transport (green squares) and reverse transport (red circles), showing hysteresis that is used to measure the maximum static friction force $F_s$. The stages of the stick-slip process (#1)-(#6) correspond to the illustrations in Fig.1B. The bold data points indicate the ion's position before a slip and only those data are used to reconstruct the force-displacement curve. **(B)** The force-displacement hysteresis loop encloses an area equal to twice the dissipated energy per slip $\Delta W$. The unit $m\omega_0^2 a$ of the applied force corresponds to $2.8\times10^{-19}$ N; here $\omega_0 = 2\pi\times364$ kHz. **(C)** The static friction force disappears for corrugations $\eta < 1$ and increases linearly with corrugation for $\eta > 1$, in excellent agreement with the Prandtl-Tomlinson model with no free parameters (red solid line). In (A), error bars indicate 1 standard deviation, while for (B) and (C) statistical error bars are smaller than the symbols. The data in (A) and (B) were measured at $\eta = 2.8$.

Here, following recent proposals (*19–22*), and enabled by the recent trapping of an ion in an optical lattice (*23–25*), we introduce an experimental system that allows us to study and control nanofriction at the individual-atom level. We form a nanofriction interface (Fig. 1A) by transporting a trapped-ion crystal with tunable spacings (*26*) over the sinusoidal potential of an optical standing wave (optical lattice), emulating an elastic crystal moving over a rigid periodic substrate. We measure the static friction force and the dissipated energy for each individual ion by tracking its position with sub-lattice-site spatial resolution, and time resolution below the thermal relaxation time scale.

$^{174}$Yb$^+$ ions, laser cooled to sub-millikelvin temperatures, are held in a linear Paul trap with harmonic confinement (*27*), where they self-organize into an inhomogeneous one-dimensional crystal owing to their mutual Coulomb repulsion. Adding the sinusoidal optical-lattice potential (*23, 28*) produces a corrugated external potential $V$ for each ion, given by $V/(m\omega_0^2 a^2) = \frac{1}{2}\left(\frac{x_i-X}{a}\right)^2 + \eta \cdot \frac{1}{4\pi^2}\cos\left(\frac{2\pi}{a}x_i\right)$ (Fig. 1A). Here $m$ is the ion's mass, $a = 185$ nm is the optical-lattice period, $x_i$ is the ion's position and $X$ is the center of the Paul trap. This potential is characterized by the dimensionless corrugation parameter $\eta$, equal to the confinement ratio $(\omega_L/\omega_0)^2$ of the lattice site vibrational frequency $\omega_L/(2\pi)$ to the Paul trap longitudinal vibrational frequency $\omega_0/(2\pi)$, both of which can be tuned over a wide range via laser intensity and static electric fields, respectively. The translation $X(t) = F(t)/(m\omega_0^2)$ of the Paul trap with respect to the optical lattice transports the ion crystal at adjustable speed, when the uniform electric force $F(t)$ is linearly ramped. The distribution of ion positions relative to the lattice can be tuned with nanometer precision via $\omega_0$, allowing us to introduce a controlled structural mismatch between object (ion crystal) and substrate (optical lattice). To remove the heat generated by friction, the ions are continuously laser cooled to temperatures much lower than the optical-lattice depth (*23*). We observe that the scattering of light by an ion is proportional to the ion's optical-lattice potential energy as a result of the lattice-assisted Raman cooling scheme (*23, 28*). Thus we can deduce the ion's position with sub-wavelength resolution during transport while its kinetic energy remains below its displacement-dependent potential energy, i.e. we can measure an ion's position before a slip, and when it has cooled down again after a slip (*28*).

We first benchmark our nanofriction simulator against the PT model by transporting a single trapped ion in the corrugated potential $V$. Under intermediate corrugation, stick-slip results from the applied-force-induced switching between the two minima of a bistable potential (Fig. 1B). As the force $F(t)$ is linearly ramped up, the ion sticks in the initial site (#1), riding up the lattice potential and increasing in fluorescence (#2), until a critical maximum static friction force $F_s$ is reached. At that point, the barrier vanishes and the initial minimum disappears, resulting in a fold catastrophe (*1*). The ion discontinuously slips from its initial site to the global minimum one site over (#3). The ion then dissipates the released energy $\Delta W$ via laser cooling, while localization in the lattice potential reduces its fluorescence again. The positions of fluorescence peaks in Fig. 2A thus correspond to the maximum static friction force $F_s$, when the ion slips. As the force ramp is reversed, hysteresis can be clearly observed in the shift $2F_s$ between the forward and reverse slips (Fig. 2A). The fluorescence increase leading up to each slip is converted to the ion's position to reconstruct the force-displacement curve enclosing the area $2\Delta W$ (Fig. 2B). We repeat the measurement at different values of the corrugation parameter $\eta = (\omega_L/\omega_0)^2$, and plot in Fig. 2C the maximum static friction force $F_s$ versus $\eta$. For $\eta < 1$ friction vanishes, as there is no bistability, and the unique potential minimum is continuously translated by the applied force. For $1 < \eta < 4.60$ the potential is bistable and $F_s$ increases with $\eta$ (linearly in the large $\eta$ limit). These results are in excellent agreement with the PT model (solid line in Fig. 2C). The regime with multiple minima $\eta > 4.60$ results in more complicated multiple-slip patterns (29), sensitive to the recooling time constant, and is not explored here.

To study multiparticle models with a trapped ion crystal, we load a desired number of ions up to $N = 6$, and control their matching to the periodic optical-lattice potential via the electrostatic harmonic confinement $\omega_0$. In the FK model, mismatch is manifested as incommensurability of the (infinite) object and substrate lattices. Although our ion crystals are finite and inhomogeneous, we find that the essence of the FK model can be captured by introducing a matching parameter $q$ that quantifies the alignment of the ions with equivalent points on the lattice when unperturbed by it. We define $q = \max_X \left[\frac{1}{N}\sum_i \sin(2\pi(x_{i0} - X)/a)\right]$, the maximum possible normalized averaged force of the optical lattice on the ions, when considering their lattice-free (unperturbed) equilibrium positions $x_{i0}$ as the harmonic trap is displaced relative to the lattice. $q$ is also related to the normalized potential barrier in the bistable energy landscape seen by the unperturbed ion crystal. By adjusting the Paul trap vibration frequency $\omega_0$, we can continuously vary the $q$ value (28) between $q = 1$, where each ion experiences an identical lattice force and the crystal behaves like a single particle (corresponding to the commensurate case in the FK model), and $q = 0$, where the lattice forces on the unperturbed crystal cancel out (analogous to an incommensurate arrangement).

For a selected matching parameter $q$, we drive the ion crystal across the lattice by linearly increasing the applied force, and measure for each ion separately the stick-slip hysteresis, extracting $F_s$ and $\Delta W$. This is performed for crystal sizes from $N = 2$ to $N = 6$ ions at a value of $\eta$ just below 4.60. As we switch from the matched case $q = 1$ to the mismatched case $q = 0$, we observe the friction change from maximal, corresponding to strong one-ion stick-slip friction for each ion, to nearly zero, corresponding to a superlubric regime, as shown in Fig. 3 for $N = 3$. Fluorescence of all three ions is plotted against the applied force in the forward and reverse directions, and the fluorescence peaks indicate the moment when each ion passes the barrier between two lattice sites. The data reveal that in the matched case, ions stick and slip together as a rigid body, with strong hysteresis between the forward and reverse transport, resulting in the maximal force-displacement hysteresis loop for each ion (middle ion shown), and maximal friction. By contrast, in the mismatched case, the ions move over the lattice in a staggered kink-like fashion, and each ion experiences almost no hysteresis or friction. Thus, the structural suppression of friction is accompanied by a transition in the nature of transport from a simultaneous slipping regime reducible to an effective single-particle PT model, to a kink propagation regime characteristic of the infinite FK model.

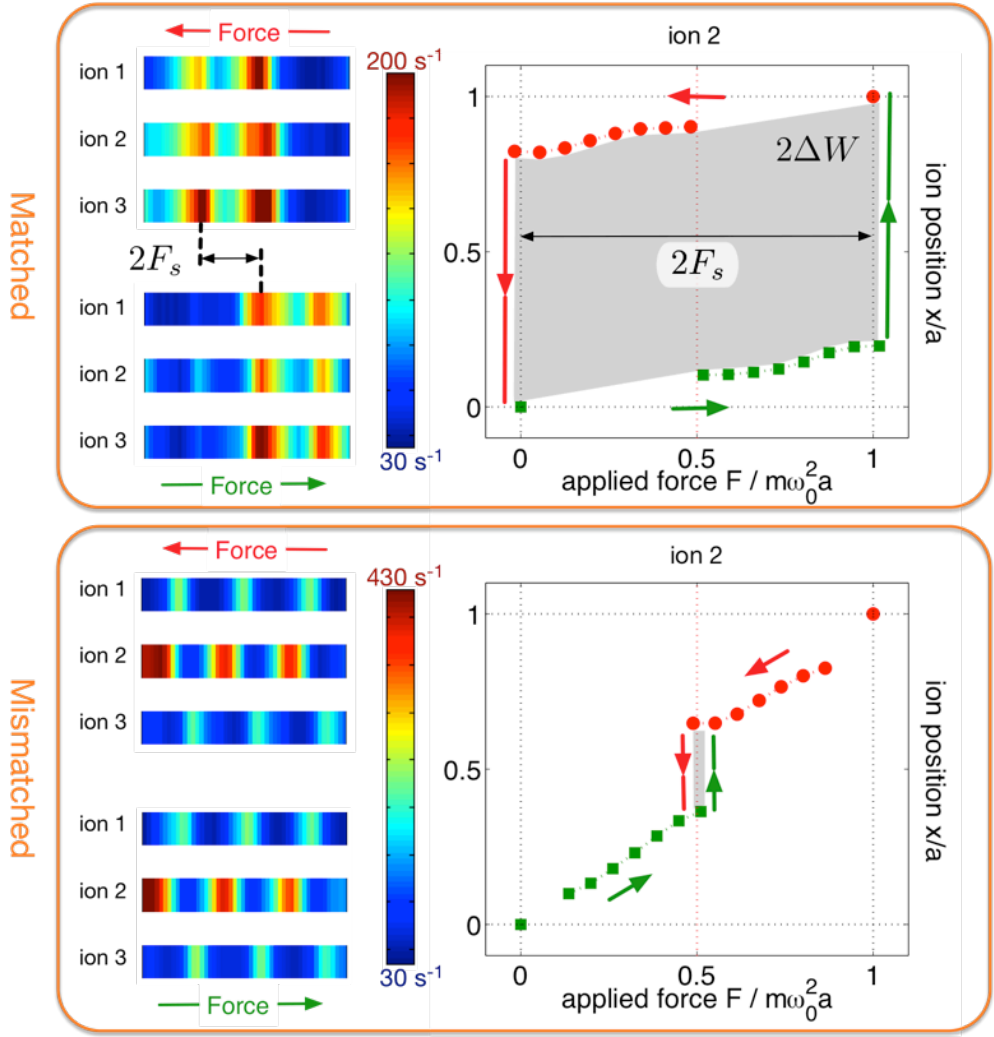

**Fig. 3. Changing friction in a 3-ion crystal from maximal to nearly frictionless (superlubric) by structural mismatch.** In the matched case (top), the ions stick and slip synchronously during transport (the observed photon detection rate for each ion, expressed in color, is maximum when the given ion slips over a potential barrier). The large hysteresis corresponds to large friction, shown here for the middle ion. In the mismatched case (bottom), the different ions slide over lattice barriers one at a time and the friction and hysteresis nearly vanish.

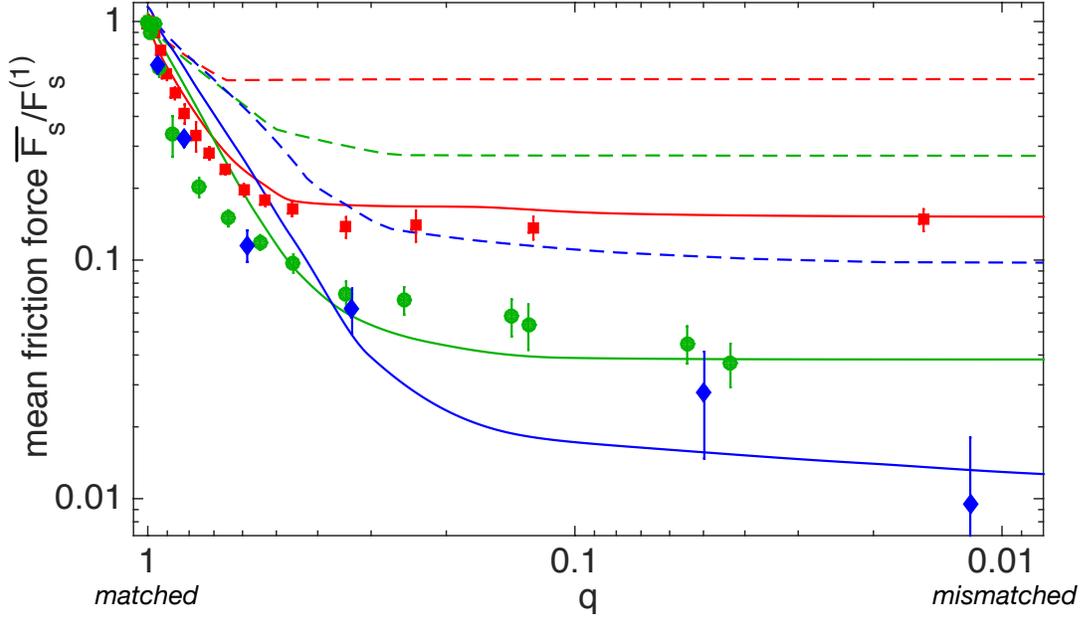

**Fig. 4. The dependence of friction on object-substrate structural matching for different crystal sizes.** Measured maximum static friction force $F_s$ for $N = 2$, 3 and 6 ions (red squares, green circles and blue diamonds, respectively), averaged over the ions and normalized to $F_s$ as measured for a single trapped ion. Error bars represent 1 standard deviation. Simulations for $N = 2$, 3 and 6 are shown for $T = 0$ (red, green and blue dashed lines, respectively) and finite $q$-dependent temperature (red, green and blue solid lines). Simulation parameters are chosen to match known experimental parameters: the measured temperature $k_B T(q = 1)/U \approx 0.05$ (corresponding to 48 μK); the optical-lattice depth $U/h = 20$ MHz (equivalent to $\eta = 4.6$); the driving velocity $v = 0.4$ mm/s; and the recooling rate constant from laser cooling $r = 2\pi \times 3$ kHz. Only the $q = 0$ temperature is fitted, yielding $k_B T(q = 0)/U = 0.15$ (corresponding to 144 μK) for all the values of $N$ shown (see (28)).

In Fig.4, we plot the measured maximum static friction force $\overline{F_s}$, averaged over the ions in the crystal, versus the matching $q$. (The dissipated energy $\overline{\Delta W}$ follows the same $q$-dependence). As $q$ is lowered from 1, the friction drops quickly, then slowly approaches a much reduced value at $q = 0$, which decreases with increasing crystal size. Notably, at $q = 0$ (mismatched limit) there is an almost tenfold reduction in friction already for $N = 2$ ions, and a hundredfold reduction for $N = 6$ ions. Numerical simulations of this behavior at zero temperature (dashed lines in Fig.4) show qualitative agreement, but fail to account for the finite temperature of the ions in the experiment. For lower $q$ values the effective barrier separating two potential minima is reduced, and the friction becomes more sensitive to temperature (28). To take temperature-induced friction reduction (thermolubricity) (1) into account, we perform full dynamics simulations accounting for the finite crystal temperature (28), and find good agreement with the experiment (solid lines in Fig.4). These simulations indicate that in the limit of low $q$, thermolubricity and superlubricity (mismatch-induced lubricity) reduce the observed friction by similar factors in our data.

Our results indicate that it may be possible to engineer nanofriction by structural control in finite-size systems. Intriguing future possibilities include the coupling to internal states of the ions (*30*) for the study of spin-dependent transport and friction (*22*), and the regime of weak periodic potentials, where quantum-mechanical tunneling may lead to new quantum phases (*19, 22*).

**Acknowledgments:** We thank Wonho Jhe and Eugene Demler for stimulating discussions, and Wonho Jhe also for critical reading of the manuscript. This work was supported by the NSF-funded Center for Ultracold Atoms (grant PHY-0551153), and Canada's NSERC Postgraduate Scholarship program. All data presented here is in the Supporting Online Material.

# Supplementary Materials

**Materials and Methods**

<u>Trapping Potentials</u>

The one-dimensional corrugated potential $V$ is produced by an intra-cavity optical standing wave superimposed on a linear Paul trap as described in detail in ref. (*23*). The TEM$_{00}$ mode of the cavity is pumped by laser light blue-detuned by 12.7 GHz from the 369.5 nm $^2S_{1/2} \to\, ^2P_{1/2}$ transition in $^{174}$Yb$^+$, resulting in optical-lattice potential minima at the nodes of the optical field. The corrugation parameter $\eta = (\omega_L / \omega_0)^2$ is defined in terms of the quantity $\omega_L = \sqrt{2\pi^2 U / (ma^2)}$, which is the oscillation frequency of an ion around an optical-lattice potential minimum in the harmonic approximation. The optical lattice coincides with the purely electrostatic axis of the Paul trap, along which the ions self-organize into a one-dimensional Coulomb crystal due to a much stronger transverse confinement by radio-frequency fields, with vibrational frequencies $\omega_{trans} / (2\pi)$ in excess of 1 MHz. Although the ions in a crystal interact via a long-range Coulomb force scaling as $|x_i - x_j|^{-2}$, it can be linearized for small ion displacements from equilibrium on the scale of the optical lattice spacing, resulting to leading order in spring forces between all ion pairs that fall off as the square of the inter-ion distance. While the Frenkel-Kontorova model assumes only nearest-neighbor spring forces, previous theoretical work (*19*) suggests that the FK model can effectively describe the long-range Coulomb system.

<u>Raman Sideband Cooling</u>

We cool the ions using a degenerate lattice-assisted Raman sideband cooling scheme *(23,31)* using the spin-1/2 Zeeman magnetic sublevels in $^{174}$Yb$^+$. The optical lattice described above, blue-detuned by 12.7 GHz, is slightly elliptically polarized. In a transverse magnetic field, the strong linearly polarized component and the weak circularly polarized component drive stimulated two-photon Raman transitions from $|n\rangle$ to $|n-1\rangle$ and $|n-2\rangle$ vibrational levels and flip the electronic spin. A circularly polarized laser beam collinear with the magnetic field and red-detuned by 100 MHz repumps the ion to the initial magnetic sublevel via a spontaneous two-photon Raman process, resulting in fluorescence which we collect. This fluorescence is modulated by a sinusoidally-varying (in space) coupling of the off-resonant, but strong $|n\rangle$ to $|n\rangle$ (carrier) stimulated Raman transitions, as well as by the sinusoidally-varying lattice-induced shift of the optical transition. This spatial variation of fluorescence is what provides us with a signal to detect an ion's position with sub-lattice-site resolution. The coupling from $|n\rangle$ to $|n-1\rangle$ and $|n-2\rangle$ also varies sinusoidally with the lattice, resulting in variations in cooling efficiency and a $q$-dependent temperature (see below).

<u>Effects of Temperature on Friction</u>

The temperature of the system can have a significant effect on stick-slip processes due to thermally induced hopping between two potential minima. In our system, the temperature is regulated by means of laser cooling, providing a dissipation mechanism to remove the heat released in the slip process. By the fluctuation-dissipation theorem, the ion follows a thermal distribution characterized by temperature $T$ and a damping rate $r$. One can define a dimensionless ratio

$\kappa = r \cdot \exp(-U/(k_B T))/(v/a)$ of the thermal hopping rate over the bare lattice potential barrier $U$ to the driven transport rate over one lattice site at velocity $v$. When $\kappa \gg 1$, stick-slip is pre-empted by thermal hopping between the sites and its effect is much reduced. To observe stick-slip in deterministic transport, thermal hopping must be negligible ($\kappa \ll 1$), which we achieve by making the temperature low $k_B T \ll U$, and by choosing the transport speed $v/a$ high enough. Since $r^{-1}$ is the characteristic time constant for recooling after a slip, the transport speed must be slower than this to avoid raising the temperature, so we operate in the regime $\exp(-U/(k_B T)) \ll (v/a)/r \ll 1$.

We measure $U/(k_B T)$ independently by the equilibrium fluorescence of the middle ion in the crystal, placed at a minimum (node) of the optical lattice. As the ion thermally samples the optical potential near the minimum, it scatters light proportional to its temperature due to the laser cooling configuration (*23*). For $N = 3$ (data presented in Fig.3), the temperature for the $q = 0$ case is measured to be at most twice the temperature for the $q = 1$ case, where $k_B T(q=1)/U \approx 0.05$. If we increase the temperature for the $q = 1$ case (by introducing excess recoil heating via slight polarization misalignment of the optical pumping beam used for the Raman cooling process (*23*)) to match that of the $q = 0$ case, we find the measured friction to be reduced only by a factor of 2, by much less than the over tenfold reduction in the measured friction when changing the mismatch from $q = 1$ to $q = 0$. From this we conclude that temperature alone cannot explain the observed reduction in stick-slip friction. We note that although this temperature measurement is insensitive to the temperature of those vibrational modes of the crystal which do not contribute to the oscillation of the middle ion (such as the stretch mode for $N = 3$), those normal modes also do not contribute to the slip of the middle ion: in the matched case, all ions slip at once, corresponding to the center-of-mass mode, while in the mismatched case, ions slip one at a time, corresponding to a localized mode, and the temperature of the relevant mode for the middle ion is measured when this ion is placed at the optical lattice minimum.

In order to further assess the effects of temperature on friction when the structural mismatch $q$ is tuned (Fig.4), we numerically simulate the full dynamics for crystals of different $N$. Using the Langevin formalism for the equations of motion subject to a fluctuating force (*1*), we obtain the $q$-dependence of friction for different temperatures. We find that in order to reproduce our experimental data, a $q$-dependent temperature is required. This is not unexpected from our cooling configuration (*23,31*), because the cooling efficiency changes depending on an ion's location in the optical potential, and $q$ tunes the arrangement of the ions relative to the optical potential. In the matched case, $k_B T(q=1)/U \approx 0.05$ is measured, while in the mismatched case, $k_B T(q=0)/U = 0.15$, as the only free parameter, produces good agreement with experimental data for $N = 2$ through 6, indicating that the effect of mismatch on temperature is insensitive to the ion number. For intermediate matching values $q$ we assume that the temperature increases linearly from $q = 1$ to $q = 0$.

We note that as the ion crystal is displaced by the applied force, because of the spatially-dependent cooling the temperature may change from the value measured with the middle ion at the optical lattice minimum. However, our simulations show that friction in the matched case $q = 1$ is highly insensitive to temperature, and we use the $q = 1$ measured temperature as input to the simulations. In the mismatched case $q = 0$ our static temperature measurement may indeed underestimate the relevant temperature in the driven situation, which may be the cause of the discrepancy between the measured $k_BT(q=0)/U \approx 0.10$ and the fitted $k_BT(q=0)/U = 0.15$ temperature values.

Controlling the q parameter

The structural mismatch parameter $q$ is measured by imaging the ion crystal as it is transported across the optical lattice in the regime where temperature dominates ($\kappa \gg 1$), and the optical-lattice potential is weak $\eta \approx 1$. In this way, the fluorescence signal from each ion reflects its average position relative to the optical-lattice potential, minimally perturbed by the optical force (aided by thermal averaging). We use these measurements to calibrate the Paul trap vibrational frequency $\omega_0$ corresponding to the minimum value of $q$, and from there on use calculated values of q corresponding to the control parameter $\omega_0$ and the given $N$. The calculated $q$ versus $\omega_0$ is periodic for $N=2$ and 3, but quasi-periodic for $N \geq 4$ (see Fig.S1) due to the inhomogeneity of the ion crystal. As a result, arbitrary tuning of $q$ with the single parameter $\omega_0$ becomes difficult with larger $N$ (the value of $q$ does not fully reach unity for $N=6$ and only gets to 0.75 for $N=10$ in the vicinity of the desired $\omega_0$, as shown in Fig.5). For future experiments with larger ion numbers, this can to some extent be rectified by controlling quartic and higher-order Paul trap potentials available in our system (27), while for very large $N$ near-square-well potentials with approximately constant ion spacings can be created by means of additional control electrodes.

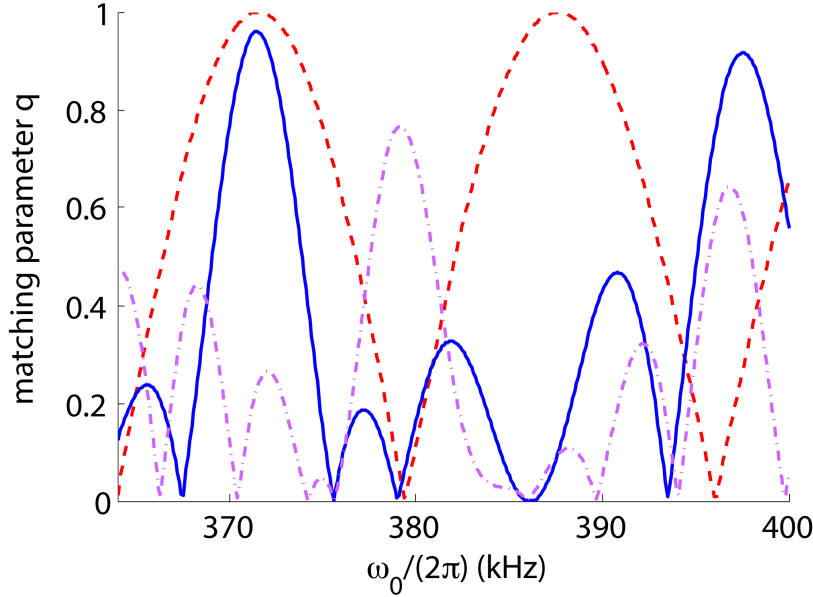

**Fig. S1.**
Calculated matching parameter $q$ as a function of the Paul trap longitudinal vibrational frequency $\omega_0$ for $N=2$ (red dashed line), $N=6$ (blue solid line) and $N=10$ (purple dot-dashed line).